\documentclass{article}

\usepackage{PRIMEarxiv}

\usepackage[utf8]{inputenc} 
\usepackage[T1]{fontenc}    
\usepackage{hyperref}       
\usepackage{url}            
\usepackage{booktabs}       
\usepackage{amsfonts}       
\usepackage{nicefrac}       
\usepackage{microtype}      
\usepackage{lipsum}
\usepackage{fancyhdr}       
\usepackage{graphicx}       
\usepackage{amsmath,amssymb,amsfonts}
\usepackage{enumitem}
\usepackage{booktabs} 
\usepackage{tabularx} 
\usepackage{float}
\title{Automated flood detection from Sentinel-1 GRD time series using Bayesian analysis for change point problems 
}

\author{
  Narumasa Tsutsumida$^{1,*}$, Tomohiro Tanaka$^{1}$, Nifat Sultana$^{1}$ \\
  $^{1}$ Graduate school of Science \& Engineering \\
  Saitama University, Japan \\
  \texttt{$^{*}$rsnaru.jp@gmail.com} \\
}

\begin{document}
\maketitle

\begin{abstract}

Current Synthetic Aperture Radar (SAR)-based flood detection methods face critical limitations that hinder operational deployment. Supervised learning approaches require extensive labeled training data, exhibit poor geographical transferability, and may fail to adapt to new regions without additional training examples. Existing approaches do not fully exploit the rich temporal information available in SAR time series, instead relying on simple change detection between pre- and post-flood images or supplementary datasets that often introduce error propagation. These limitations prevent effective automated flood monitoring in data-scarce regions where disaster response is most needed. To address these limitations, we develop a novel training-free approach by adapting Bayesian analysis for change point problems, specifically for automated flood detection from Sentinel-1 Ground Range Detected time series data. Our method statistically models the temporal behavior of SAR backscatter intensity over a one-year baseline period, then computes the posterior probability of change points at flood observation dates. This approach eliminates supervised learning dependencies by using Bayesian inference to identify when backscatter deviations exceed expected normal variations, leveraging inherent statistical properties of time series data. Validation across three diverse geographical contexts using the UrbanSARFloods benchmark dataset demonstrates superior performance compared to conventional thresholding and deep learning approaches, achieving F1 scores up to 0.75. This enables immediate deployment to any region with SAR coverage, providing critical advantages for disaster response.

\end{abstract}



\section{Introduction}
\label{sec:Introduction}

Flooding is one of the most frequent and devastating natural disasters globally \cite{Alfieri2017-ma,Najibi2018-ta}, resulting in substantial loss of life, damage to infrastructure, and severe economic consequences \cite{world2021wmo, Kundzewicz2014-ge,Tellman2021-hv,Jonkman2024-hi, Rentschler2021-dc}. The economic impact and human toll of flooding necessitate effective monitoring systems for early detection and rapid response \cite{Oddo2019-or,Rentschler2020-pm}. While satellite remote sensing has emerged as a critical tool for flood mapping \cite{Oddo2019-or, Katiyar2021-fy, Shen2019-aj,Tay2020-cp}, providing situational awareness to disaster responders and enabling damage assessments \cite{Browder2021-io},  the widespread adoption of operational flood mapping remains limited due to fundamental methodological challenges in remote sensing data analytics.

Synthetic Aperture Radar (SAR) offers significant advantages for flood monitoring through its ability to operate in the microwave portion of the electromagnetic spectrum, enabling all-weather, day-night imaging capability regardless of cloud cover. The Sentinel-1 mission by the European Space Agency (ESA) has greatly enhanced the availability of SAR data, providing global coverage with a revisit time of 6-12 days, free data access, high spatial resolution (5-20 m), and wide swath width (250 km), making it well-suited for mapping flood extents at regional to continental scales \cite{Schlaffer2015-uv}. While Sentinel-1 data is available in both Single Look Complex (SLC) and Ground Range Detected (GRD) formats, most operational applications utilize GRD products due to their reduced preprocessing requirements and accessibility for operational monitoring systems.

Despite these advantages, current SAR-based flood mapping approaches face significant methodological limitations. Many existing techniques rely on supervised classification methods that require extensive training data \cite{Shastry2023-wl,Wu2023-kc,Ghosh2024-ub}, creating substantial barriers to operational flood monitoring in data-scarce regions. The transferability of trained models to new flood scenarios or geographical contexts is often limited \cite{Zhao2024-lv, Zhao2025-cx}, necessitating additional data collection and model retraining for each new application area. 
A further critical limitation is the failure to fully exploit the rich temporal information available in SAR time series. 
Current approaches frequently depend on supplementary datasets such as optical satellite imagery \cite{DeVries2020-ga,Farhadi2025-bs}, digital elevation models \cite{Martinis2018-pi}, or hydrological information, creating multi-source integration challenges that increase computational overhead, introduce error propagation, and compromise the temporal consistency needed for real-time flood monitoring.

To address these limitations, some studies have focused on developing analytical frameworks that fully exploit time-series approaches for Sentinel-1 GRD-based flood monitoring.
However, these methods remain methodologically constrained, relying primarily on simple change detection between pre- and post-flood images or on conventional statistical measures that inadequately capture the temporal dynamics of backscatter behavior. Traditional change detection methods struggle to distinguish flood-induced backscatter changes from natural seasonal variations or other land surface modifications (e.g., agricultural practices, vegetation phenology), often resulting in high false alarm rates. Furthermore, these methods typically require manual threshold selection and extensive parameter tuning, hindering their automation potential.
While Bayesian time series analysis methods have shown promise for detecting abrupt changes \cite{Barry1993-qp,Erdman2008-kv}, they have not been effectively adapted to exploit the unique temporal characteristics of SAR backscatter for flood detection. This represents a significant research gap, as SAR time series inherently contain the temporal context necessary to distinguish flood events from background variability without requiring external data sources—offering the potential for fully automated flood monitoring systems.

This study aims to overcome these fundamental methodological challenges by developing an automated solution for flood detection using Sentinel-1 GRD time series data. We propose a pixel-wise Bayesian statistical model for SAR-based flood mapping that operates without the need for training data or pre-existing water masks.  Unlike existing approaches, our method leverages a Bayesian analysis of Change Point problems (BCP) as proposed by Barry and Hartigan \cite{Barry1993-qp} to identify flood-induced changes in Sentinel-1 GRD time series data. Our workflow distinguishes between permanent water bodies and temporary flood inundations without requiring ancillary data by leveraging the temporal change characteristics in the probability domain rather than the backscatter domain. By working directly with readily available GRD products, our approach eliminates the complex preprocessing requirements associated with SLC data while still achieving robust flood detection performance. By addressing key methodological limitations in current flood detection approaches, our BCP-based technique offers an advancement in the field of SAR-based flood monitoring.


\section{Related works}
\label{sec:RelatedWorks}


\subsection{Flood monitoring by remote sensing data}

While optical satellite platforms such as Sentinel-2 and PlanetScope offer high spatial resolution and multispectral capabilities for flood mapping \cite{Portales-Julia2023-pz, Wieland2019-je,Farhadi2024-bm}, cloud cover significantly impedes continuous monitoring during critical initial flood phases \cite{ Farhadi2024-bm}. 
SAR presents a compelling alternative due to its all-weather imaging capability \cite{Landuyt2019-cz,Amitrano2018-ie, Amitrano2024-av, Shen2019-aj}. SAR discriminates between water and non-water surfaces based on distinctive backscattering properties, with smooth water surfaces typically appearing dark due to specular reflection.
Interferometric SAR (InSAR) utilizes phase coherence information between acquisitions to detect inundation patterns \cite{garg2024unlocking}, demonstrating effectiveness in complex urban environments \cite{Chini2019-me,Pelich2022-rk}. However, the practical application of InSAR for timely flood monitoring is hindered by significant challenges. Normally, the extensive processing needed to derive these parameters often prevents real-time flood monitoring, limiting the method's operational utility in urgent disaster response scenarios.

Deep learning methods have demonstrated exceptional capabilities for SAR-based flood detection \cite{Saleh2024-vd, Bereczky2022-qy}, with recent studies reporting classification accuracies exceeding 98\% in certain contexts \cite{Ghosh2024-ub}. Specialized datasets like Sen1Floods11 \cite{Bonafilia2020-sk}, UrbanSARFloods \cite{Zhao2024-lv}, and Kuro Siwo \cite{Bountos2023-jg} have facilitated model training across diverse geographical contexts.
Nevertheless, significant challenges persist from SAR time series: (1) extensive labeled training data requirements, (2) limited model transferability across geographical contexts \cite{Wu2023-kc}, (3) difficulty distinguishing between permanent water bodies and temporary inundation \cite{Notti2018-vs, Farhadi2025-bs}, and (4) complex backscatter interpretation in urban or vegetated areas \cite{Amitrano2024-av}.

\subsection{SAR Time Series for Flood Detection}

Recent research has increasingly leveraged the temporal dimension of SAR data for improved flood detection \cite{Schlaffer2015-uv}. Time series approaches exploit the sequential nature of acquisitions to identify anomalous patterns indicative of flood events, particularly effective with Sentinel-1's consistent 6-12 day revisit cycle \cite{Oddo2019-or}.
Schlaffer et al. (2015) proposed a harmonic analysis approach using multi-temporal SAR data to detect floods based on deviations from expected seasonal patterns \cite{Schlaffer2015-uv}. Similarly, Tsyganskaya et al. (2018) \cite{Tsyganskaya2018-ts} and Tsyganskaya et al. (2019) \cite{Tsyganskaya2019-nv} employed time series analysis for wetland change detection and flooded vegetation mapping using Sentinel-1 data. These methods enable more robust discrimination between permanent water bodies and temporary inundation.

Conventional Bayesian methods offer advantages for hydrological applications by incorporating prior knowledge and quantifying uncertainty \cite{Barry1993-qp, Erdman2008-kv}. D'Addabbo et al. (2016) \cite{D-Addabbo2016-yi} developed a Bayesian network model fusing SAR and ancillary data, and Giordano et al. (2023) \cite{Giordano2023-ys} applied Bayesian statistical modeling to multi-temporal Sentinel-1 stacks for high-resolution flood monitoring.
However, existing Bayesian approaches primarily focus on retrospective analysis rather than up-to-date detection of the latest observation. Most methods lack specific adaptations for the characteristic SAR backscatter patterns during flood events. These patterns typically include sharp decreases during inundation followed by gradual increases during recession, which limits their effectiveness for timely flood monitoring.

\subsection{Research Gap and Our Contribution}

Previous studies face challenges in SAR-based flood monitoring that our approach addresses: (1) extensive training data requirements limiting applicability in flood detection, (2) domain adaptation difficulties across flood scenarios, and (3) challenges distinguishing permanent water bodies from flood inundation.
Our approach with BCP operates without training data, processes Sentinel-1 GRD time series directly, and leverages temporal patterns to differentiate water classes without pre-existing water masks. This approach fills a critical gap in flood monitoring studies, offering a complementary method particularly valuable when timely detection is essential for emergency response.

\section{Methods}
\label{sec:Methods}

\begin{figure*}[!t]
    \centering
    \includegraphics[width=0.85\textwidth]{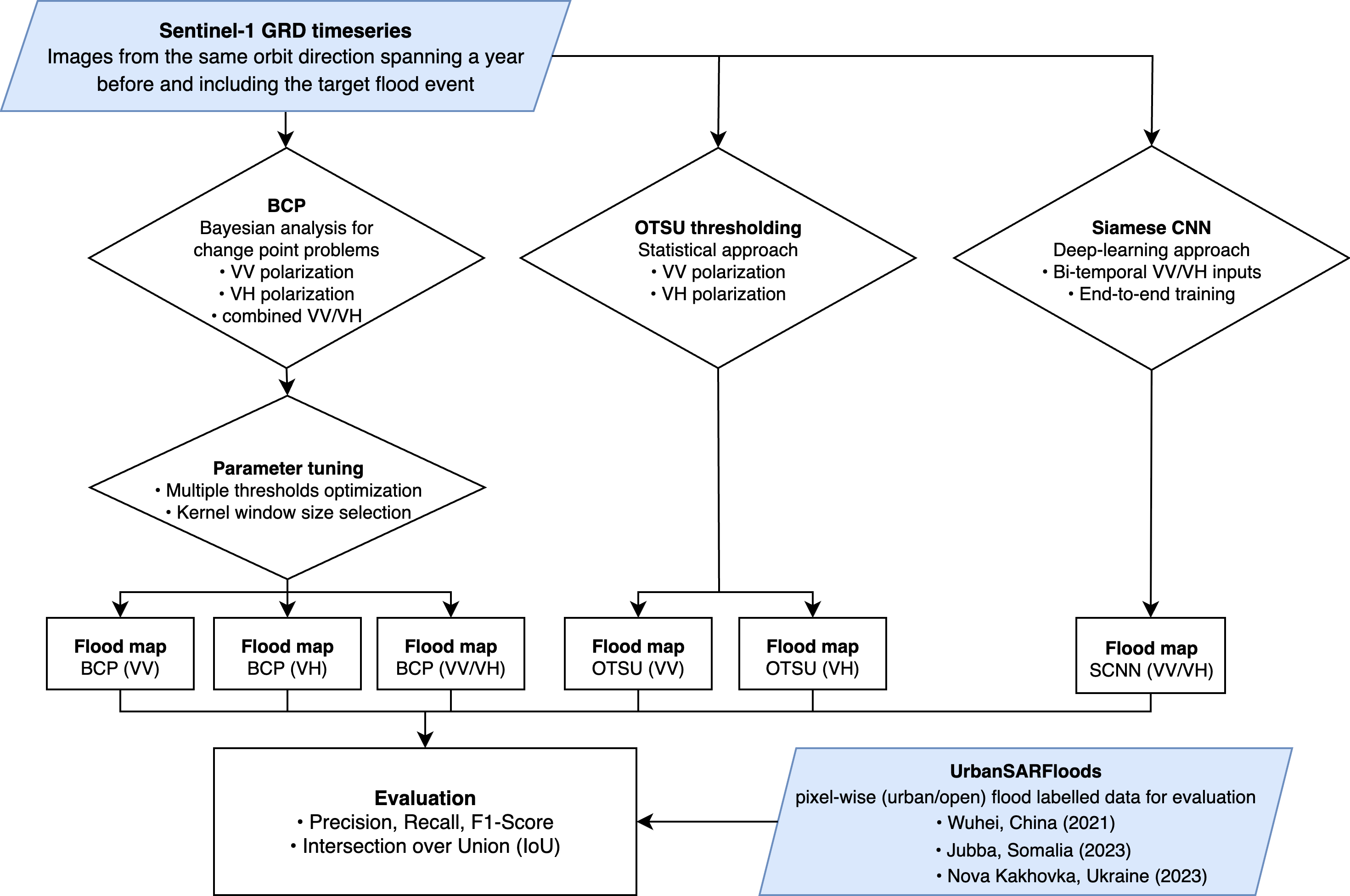}
    \caption{Methodological flowchart comparing three flood detection methods: Bayesian analysis of change point problems (BCP), Otsu's thresholding, and Siamese CNN. Blue boxes indicate external datasets, including input Sentinel-1 data and the UrbanSARfloods evaluation dataset. }
    \label{fig:flowchart}
\end{figure*}

The workflow of our approach consists of three main components: (1) reference data and Sentinel-1 GRD time series data preparation, (2) BCP analysis, and (3) post-processing with spatial filtering (Figure \ref{fig:flowchart}). The workflow is designed to process GRD data directly and provide automated flood extent mapping without requiring training data or extensive manual intervention.

\subsection{Data Preparation}
\label{sec:data_prep}
\subsubsection{UrbanSARFloods}
As reference data, we utilized the UrbanSARFloods dataset \cite{Zhao2024-lv}, a comprehensive benchmark for flood detection using Sentinel-1 SAR data. This dataset encompasses 8,879 non-overlapping image chips of 512×512 pixels at 20-m spatial resolution, covering an area of 807,500 km\textsuperscript{2} across 18 flood events on 5 continents. Every pixel in this dataset is classified into three distinct categories: non-flooded areas, flooded open areas, and flooded urban areas. UrbanSARFloods is systematically divided into training (4,501 chips), validation (1,970 chips), and testing (2,408 chips) data. The testing data are independent of flood events in the training and validation data. This dataset ensures its quality by a two-tier annotation process, which combines semi-automatic labeling through conventional remote sensing techniques with manual verification using high-resolution optical data from PlanetScope and UAV imagery \cite{Zhao2024-lv}. While the BCP analysis does not require any training/validation data, we used them for the comparative analysis described in Section \ref{sec:eval}. 

\begin{figure*}[!t]
    \centering
    \includegraphics[width=0.85\textwidth]{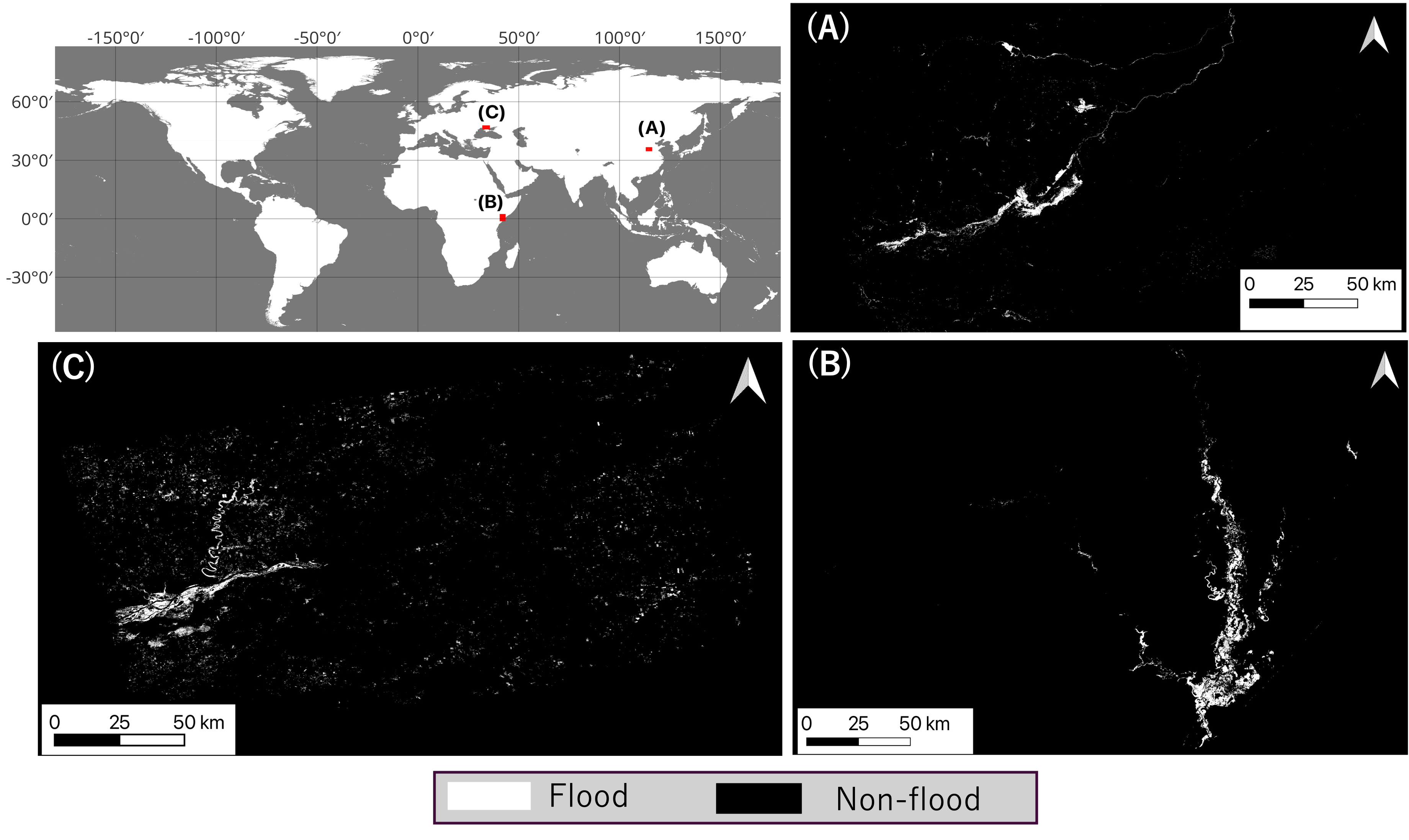}
    \caption{Overview map of tested flood events for (A) Weihui, China, (B) Jubba, Somalia, and (C) Nova Kakhovka, Ukraine. }
    \label{fig:overviewmap}
\end{figure*}

To evaluate the geographical robustness of our method, we conducted extensive testing across three diverse test sites found in UrbanSARFloods: Weihui (China), Jubba (Somalia), and Nova Kakhovka (Ukraine) (Figure \ref{fig:overviewmap}). 
Weihui represents an urban environment with complex infrastructure and dense building patterns. In such settings, SAR backscatter is influenced by double-bounce reflection mechanisms between water surfaces and vertical structures, creating complex scattering patterns that can confound conventional flood detection approaches. The regular grid pattern of roads and buildings creates distinctive backscatter signatures that require careful interpretation.
Jubba exemplifies a rural, agricultural landscape with varying topography and seasonal vegetation patterns. In these areas, the temporal variability of vegetation growth can mask or mimic flood signatures, particularly in partially submerged vegetation where volume scattering mechanisms dominate. The predominantly open terrain presents different challenges compared to urban environments, with larger homogeneous regions but greater seasonal variability.
Nova Kakhovka represents a mixed environment with both urban elements and natural landscapes, including significant riverine features. This combination of built environment and natural terrain presents a particularly challenging test case, as algorithms must correctly distinguish permanent water bodies (river) from newly flooded areas while accounting for different backscatter behaviors across varied land cover types.

\subsubsection{Sentinel-1 GRD}
Sentinel-1 GRD offers VV and VH polarization modes that measure radar backscatter amplitude using C-band SAR instruments. The data undergoes multi-looking processing and is projected onto the WGS84 ellipsoid, converting radar measurements into ground coordinates for geographic analysis. This multi-looking process reduces noise while preserving backscatter amplitude, though it removes phase information.
For each flood event, we collected time series data spanning one year before the event using Google Earth Engine, to establish a robust baseline for change detection. 
We didn't take into account of further processing such as speckle filtering and radiometric terrain normalization to ensure the sensitivity of time-series variability of backscatter intensity to changes in surface properties before applying the BCP. 
To account for differences in viewing geometry and illumination conditions over time, we filtered the timeseries data in the platform type (A or B) and the orbit direction (ascending or descending) that influences backscatter response due to varying incidence angles and shadow effects based on the observation at the flood event (Table \ref{tab01}).

While Sentinel-1 GRD data is originally available at 10-m spatial resolution, we deliberately aggregated it to 20-m resolution to match the UrbanSARFloods dataset. This methodological decision was critical because aggregating results from 10-m to 20-m after processing would potentially mask pixel-level misclassifications at the original resolution. By performing the aggregation before analysis, we ensured a more accurate and fair comparison between our predictions and the reference data.

\begin{table*}[!t]
\centering
\caption{Tested flood events and observation conditions during the flood event.}
{\small
\begin{tabular}{lrrrr} \hline
   Selected event & Country & Orbit & Day of image for annotation \\ \hline
   Weihui &  China & Ascending  &  27th July, 2021  \\
   Jubba & Somalia & Decending  &  1st December, 2023 \\
   Nova Kakhovka & Ukraine & Ascending & 9th June, 2023 \\ \hline
\end{tabular}
}
\label{tab01}
\end{table*}

\subsection{Bayesian analysis for change point problems}
\label{sec:bcp}

The BCP assumes the observations $X_1, ..., X_n$ are independent and normally distributed with $X_i \sim N(\mu_i, \sigma^2)$. The probability of a changepoint at position $i$ is denoted by $p$ and is assumed to be independent at each $i$. The prior distribution of the mean $\mu_{ij}$ for the block from position $i+1$ to $j$ is chosen to be $N(\mu_0, \sigma_0^2/(j-i))$. This prior allows the detection of subtle changes given sufficient data.

The BCP uses a partition $\rho = (U_1, ..., U_n)$, where $U_i = 1$ indicates a changepoint at position $i+1$. At each step of the Markov chain Monte Carlo (MCMC) procedure, a value of $U_i$ is drawn from the conditional distribution of $U_i$ given the data and current partition. The transition probability $p$ for the conditional probability of a changepoint at position $i+1$ is calculated as:

\begin{equation}
\begin{split}
\frac{p_i}{1-p_i} &= \frac{P(U_i=1 | \mathbf{X}, U_j, j \neq i)}{P(U_i=0 | \mathbf{X}, U_j, j \neq i)} \\
&= \frac{\left[\int_{0}^{\gamma} p^{b}(1-p)^{n-b-1} dp\right]}{\left[\int_{0}^{\gamma} p^{b-1}(1-p)^{n-b} dp\right]} \cdot 
\frac{\left[\int_{0}^{\lambda} \frac{w^{b/2}}{(W_1+B_1 w)^{(n-1)/2}} dw\right]}{\left[\int_{0}^{\lambda} \frac{w^{(b-1)/2}}{(W_0+B_0 w)^{(n-1)/2}} dw\right]}
\end{split}
\end{equation}

where $b$ is the number of blocks if $U_i=0$, $W_0$ and $B_0$ are the within and between block sums of squares when $U_i=0$, $W_1$ and $B_1$ are the equivalent when $U_i=1$, and $\gamma$ and $\lambda$ are tuning parameters.

The within block sum of squares $W$ is calculated as:
\begin{equation}
W = \sum_{ij \in \rho} \sum_{l=i+1}^{j}(X_l - \bar{X}_{ij})^2
\end{equation}
where $\bar{X}_{ij}$ is the mean of the observations in the block from $i+1$ to $j$.

The between block sum of squares $B$ is calculated as:
\begin{equation}
B = \sum_{ij \in \rho} (j-i)(\bar{X}_{ij} - \bar{X})^2
\end{equation}
where $\bar{X}$ is the overall mean of the data.

We implemented BCP using the bcp R package \cite{Erdman2008-kv}.
We selected the default tuning parameters of $\gamma=\lambda=0.2$ as recommended by Barry and Hartigan \cite{Barry1993-qp, Erdman2008-kv}. 
Posterior means and probabilities were calculated based on 500 MCMC iterations after discarding the first 50 as burn-in, a configuration that achieved convergence while maintaining computational efficiency.

To our SAR time series data over a 1-year period prior to each flood event, we use the posterior probability of a changepoint at the last observation during the flood as input for the subsequent processing. This is expected to detect potential changes in the SAR signal associated with the flood while accounting for normal seasonal variations in backscatter.
Three different sets of inputs to the BCP analysis were considered to evaluate the sensitivity and performance of our approach: (1) VV polarization only, (2) VH polarization only, and (3) VV-VH combined. The VV polarization is generally more sensitive to surface roughness and typically provides stronger contrast between water and non-water surfaces. The VH cross-polarization is often more sensitive to volume scattering and can provide complementary information, particularly in vegetated areas. The combined VV-VH approach leverages information from both polarizations, potentially enhancing the detection capabilities across diverse environmental contexts. By examining these three input configurations, we assessed which combination of polarimetric information most effectively captured flood-induced changes in the SAR backscatter time series.

\subsection{Post-processing}
\label{sec:postprocessing}
Speckle in SAR imagery results from the coherent summation of electromagnetic waves scattered from multiple targets within each resolution cell, creating a characteristic granular pattern that is an inherent property of coherent imaging systems. We applied the BCP to a Sentinel-1 GRD time series that inherently exhibits speckle patterns due to coherent imaging, which may lead to false detections of flooded areas. To mitigate speckle while preserving spatial details, we applied a moving window kernel \cite{Fauvel2012-au,Qiu2004-cl} to the results of the BCP analysis.
The kernel that calculated the average of $w \times w$ pixel values centered on the target observation was applied to the BCP output. The window size $w$ is a parameter that can be adjusted based on the spatial resolution of the input imagery and the desired level of smoothing. After applying the filter, the smoothed values were thresholded using a parameter $t$. Pixels with values above this threshold were considered as potential flood areas, while those below were masked out.

To determine optimal parameters for our flood detection method, we conducted a systematic parameter sensitivity analysis focusing on the $w$ and $t$ as post-processing steps.
For the window size parameter, we explored values ranging from 3 to 15 pixels with a step size of 2. The lower bound of 3 represents the minimum window size needed for effective speckle mitigation while the upper bound of 15 was chosen to avoid excessive smoothing that could eliminate small but significant flooded areas. The 20 m spatial resolution of our analysis data meant that a window size of 3 would cover an area of 60 m × 60 m on the ground, while a window size of 15 would cover 300 m × 300 m. 
For the threshold parameter $t$, we investigated values between 0.1 and 0.9 with a step size of 0.1. This comprehensive range covers the entire probability space from very permissive detection (0.1) to highly selective identification (0.9). Theoretically, changepoint probabilities above 0.5 indicate a more likely than not change, but in practice, lower thresholds may be effective for detecting subtle hydrological changes.
The parameter selection directly influences accuracy assessment metrics through precision-recall trade-offs. Higher threshold values reduce both true and false positives, generally increasing precision but decreasing recall as actual flood areas may be missed. Larger window sizes reduce false detection through spatial smoothing, typically improving precision, though recall may decrease if small legitimate flood areas are also smoothed out. Understanding these relationships is crucial for interpreting the parameter optimization results and ensuring transparency in the accuracy assessment process.
Thus, the parameter optimization process involved evaluating all 63 combinations of window sizes and thresholds (7 window sizes × 9 threshold values) across multiple flood events in diverse geographical settings to ensure the robustness and transferability of our approach.

\section{Evaluation}
\label{sec:eval}

To demonstrate the effectiveness of our proposed approach, we compared our results with both conventional and modern deep-learning approaches for flood detection.

\subsubsection{Otsu's Thresholding Method}
As a conventional baseline, we implemented the Otsu's threshold algorithm \cite{Otsu1979-dj}. This non-parametric technique determines an optimal threshold by maximizing the between-class variance between flood and non-flood pixels in SAR imagery. We applied the Otsu's thresholding approach to two different input configurations: VV polarization only and VH polarization only.

The Otsu's threshold algorithm operates by computing the weighted sum of variances of the two classes:
\begin{equation}
\sigma_{w}^{2}(t)=\omega_{0}(t)\sigma_{0}^{2}(t)+\omega_{1}(t)\sigma_{1}^{2}(t)
\end{equation}
where $\omega_{0}$ and $\omega_{1}$ represent the class probabilities, $\sigma_{0}^{2}$ and $\sigma_{1}^{2}$ denote the class variances, and $t$ is the threshold value.
We followed standard practices in remote sensing literature for Otsu's thresholding approach, including Lee filter application for speckle reduction, followed by histogram equalization preprocessing to enhance contrast in the SAR imagery before applying the threshold calculation.

\subsubsection{Siamese CNN Method}
As a representative of modern deep learning approaches, we built a Siamese Convolutional Neural Network (CNN) from the training and validation sets of the UrbanSARFloods dataset. The network architecture consists of dual processing branches with shared weights to compare the latest observation with the second latest observation.
The model takes two input images of size 512×512 pixels with 2 channels each (VV and VH polarizations): Input A (latest observation) and Input B (second latest observation). Each image is processed through identical convolutional branches with shared weights, ensuring that features are extracted using the same transformations. The encoder pathway implements sequential convolutional blocks with increasing filter depths (32, 64, and 128 filters), where each block contains:
\begin{itemize}
\item  3×3 convolution with 'same' padding to maintain spatial dimensions  
\item  Batch normalization to stabilize and accelerate training   
\item  ReLU activation function to introduce non-linearity  
\end{itemize}
The decoder pathway incorporates skip connections from corresponding encoder layers to preserve spatial details through addition operations rather than concatenation. A custom subtraction layer computes the difference between the processed features from both branches to identify changed regions, followed by a 1×1 convolution with sigmoid activation to produce the final flood probability map.

The model was trained using binary cross-entropy loss function, with consideration for class imbalance where the positive class weight was calculated to be approximately 35.67. We used the Adam optimizer with an initial learning rate of 0.001 and gradient clipping ($clipnorm=1.0$), along with a ReduceLROnPlateau scheduler that reduced the learning rate by half after 5 epochs of no improvement. The model was trained for up to 200 epochs with early stopping ($patience=20$) based on validation loss.
Each input was normalized using Z-score normalization and clipped to ±3 standard deviations to handle outliers. NaN values were replaced with zeros to ensure numerical stability.

\subsubsection{Accuracy assessment}

We evaluated our method's performance using standard metrics including precision, recall, F1 score, and Intersection over Union (IoU). The evaluation was conducted in two ways: first as an overall flood detection assessment, and then separately for urban and open areas to account for different challenges in each context. 
For the overall flood detection evaluation, we considered both flood open and urban pixels as positive cases. For the separate analysis, we evaluated the performance independently for each land cover type according to the reference data.

Precision represents the proportion of correctly predicted flood water pixels among all positive predictions, while recall measures the proportion of actual flood water pixels that were correctly identified. Precision and recall are calculated as follows:
\begin{equation}
Precision = \frac{N_{TP}}{N_{TP} + N_{FP}}
\end{equation}
\begin{equation}
Recall = \frac{N_{TP}}{N_{TP} + N_{FN}},
\end{equation}
where $N_{TP}$ represents the number of true positives (correctly predicted flood water pixels), $N_{FP}$ represents the number of false positives (incorrectly predicted as flood water), and $N_{FN}$ represents the number of false negatives (flood water pixels not detected).
The F1 score, which provides a balanced measure of the model's performance, is calculated as:
\begin{equation}
F1 = 2 \times \frac{Precision \times Recall}{Precision + Recall}
\end{equation}

The IoU was aimed to quantify the spatial overlap between the predicted flood water regions and the ground truth annotations:
\begin{equation}
IoU = \frac{Area\ of\ Overlap}{Area\ of\ Union} = \frac{N_{TP}}{N_{TP} + N_{FP} + N_{FN}}
\end{equation}

For both the overall and separate evaluations, these metrics were calculated for each test site (Weihui, Jubba, and Nova Kakhovka).

\section{Results}
\label{sec:Results}

\subsection{Original BCP Outputs}
The BCP analysis outputs revealed that the estimated flood extents are sensitive to the input variables (Figure \ref{fig:bcp_outputs}).
The visual comparison of the BCP outputs suggests that the BCP results using VH alone seem less capable of detecting flood inundation than the results using VV or VV-VH input. The use of VV-VH input delineates the flood extent more clearly than the others at all three sites.

\begin{figure*}[!t]
    \centering
    \includegraphics[width=0.85\textwidth]{Figures/Figure3_python.png}
    \caption{Randomly selected patches for: Jubba (Somalia) (a) Bing map that represent non-flood condition, (b) Referenced UrbanSARFloods data set, (c) BCP results used VV input only, (d) BCP results used VH input only, (e) BCP results used VV-VH input; Weihui (China) (f) Bing map that represent non-flood condition, (g) Binary UrbanSARFloods data set, (h) BCP results used VV input only, (i) BCP results used VH input only, (j) BCP results used VV-VH input; Nova Kakhovka (Ukraine) (k) Bing map that represent non-flood condition, (l) Binary UrbanSARFloods data set, (m) BCP results used VV input only, (n) BCP results used VH input only, (o) BCP results used VV-VH input. }
    \label{fig:bcp_outputs}
\end{figure*}

\subsection{Parameter Sensitivity Analysis}
\begin{figure*}[!t]
    \centering
    \includegraphics[width=0.85\textwidth]{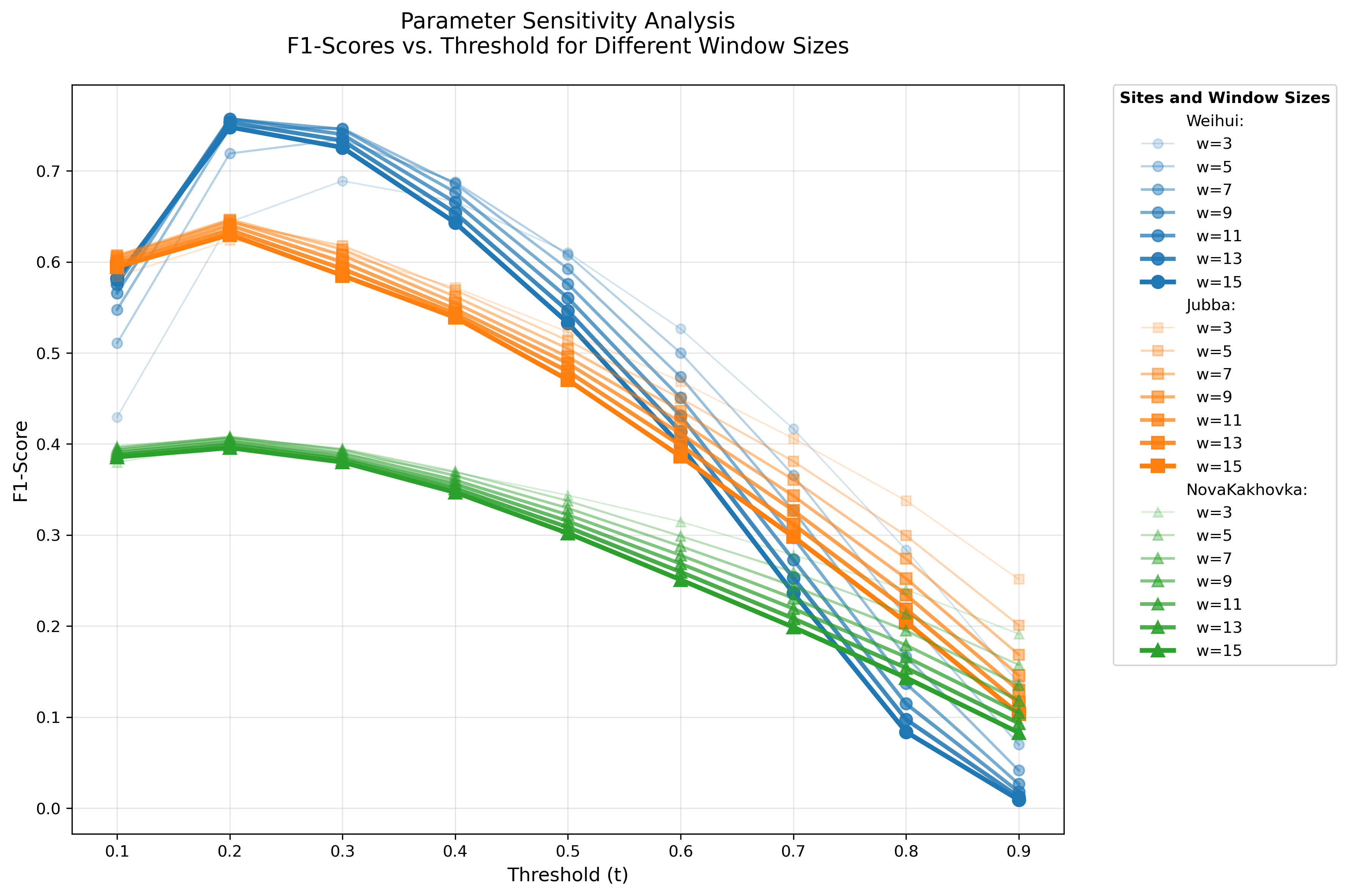}
    \caption{Parameter sensitivity analysis for post-processing of BCP flood detection results using a randomly selected patch from Jubba, Somalia. The grid shows flood detection outputs for a representative 512×512 pixel patch across different combinations of filter window sizes (w = 3, 5, 7, 9, 11, 13, 15) and probability thresholds (t = 0.1 to 0.9). Blue areas indicate detected flood pixels, white areas represent non-flooded regions. The reference ground truth is shown in the leftmost column. F1 scores are displayed for each parameter combination to demonstrate performance variations.}
    \label{fig:bcp_with_parameters}
    \end{figure*}

The original BCP results are the posterior probabilities of flood occurrence at each pixel. To delineate the flood extents, we post-processed the BCP results with different parameter settings for threshold and window size and created the binary (flood/non-flood) outputs. Considering the result using VV-VH input only, as this is thought to be the best performance for the BCP, Figure \ref{fig:bcp_with_parameters} shows the comparison of the binary outputs with different parameter settings for a selected patch of Jubba (Somalia) as an example. The larger $w$ tends to overestimate the flood extent, while the smaller $w$ may fail to mitigate speckle. The larger $t$ picks up only the flood extent with the higher confidence, while the smaller $t$ may include false positives.

\begin{figure*}[!t]
    \centering
    \includegraphics[width=0.85\linewidth]{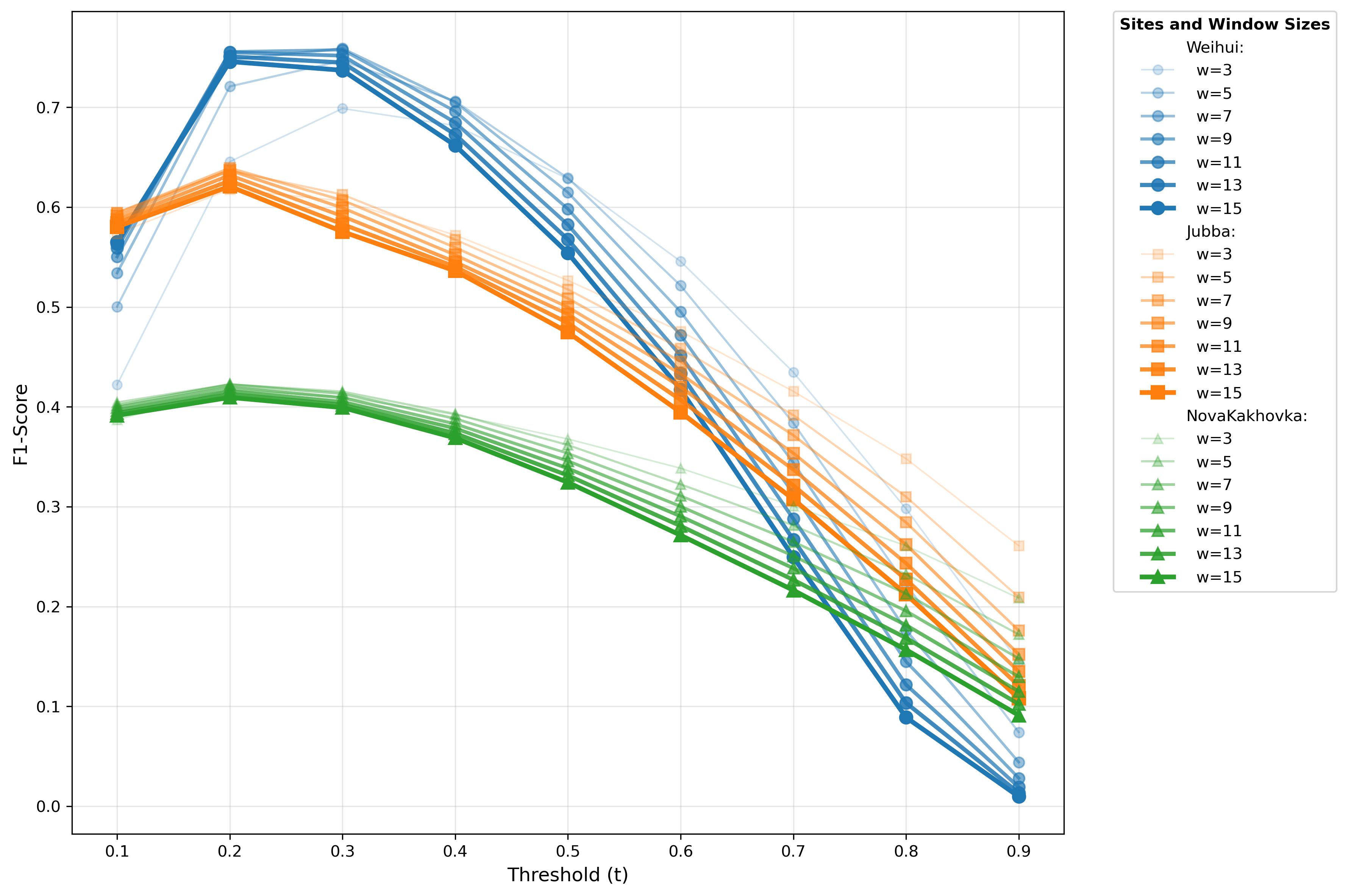}
    \caption{Parameter sensitivity analysis of flood detection method based on F-1 score against threshold and window size in Weihui (China), Jubba (Somalia), and Nova Kakhovka (Ukraine). }
    \label{fig:sensitivity}
    \end{figure*}

To evaluate the performance of the estimated flood extents, we explored the parameter sensitivity based on the F-1 score against threshold and window (Figure \ref{fig:sensitivity}).
The optimal parameter combination emerged as a threshold value of 0.2 and a window size of 9 pixels, achieving a mean F1-score of 0.60 across all test sites. However, performance varied substantially across locations, with the highest performance at Weihui (F1-score: 0.76), followed by moderate performance at Jubba (0.65), and lower performance at Nova Kakhovka (0.41).

We identified 44 of 63 parameter combinations that showed no statistically significant difference from the optimal configuration ($p > 0.05$). These combinations formed several key clusters: combinations with $t=0.2$ and window sizes 7, 11, and 13 (Mean F1 $>$ 0.59, see table \ref{tab:top_combinations}); combinations with $t=0.3$ and window sizes 5-15 (Mean F1 $>$ 0.56); and some combinations with $t=0.1$ (Mean F1 $>$ 0.50).
Performance degradation became statistically significant for threshold values $\geq$ 0.4, particularly when combined with extreme window sizes (Figure \ref{fig:sensitivity}). The decline in performance showed a consistent pattern with increasing threshold values and window size.

\begin{table*}[!t]
\centering
\caption{Top performing parameter combinations and their statistical significance against the result with parameters of t=0.2 and w=9. Selected results only with t=0.2.}
\label{tab:top_combinations}
{\small
\begin{tabular}{llllll}
\hline
Threshold (t) & Window Size (w) & Mean F1-Score & Std F1-Score & p-value \\
\hline
0.2 & 9 & 0.60 & 0.18 & -  \\
0.2 & 7 & 0.60 & 0.18 & 0.69 \\
0.2 & 11 & 0.60 & 0.18 & 0.22 \\
0.2 & 13 & 0.60 & 0.18 & 0.06 \\
0.2 & 5 & 0.59 & 0.16 & 0.43 \\
\hline
\end{tabular}
}
\end{table*}

\subsection{Visual Analysis Across Sites}

\begin{figure*}[!t]
\centering
\includegraphics[height=0.9\textheight]{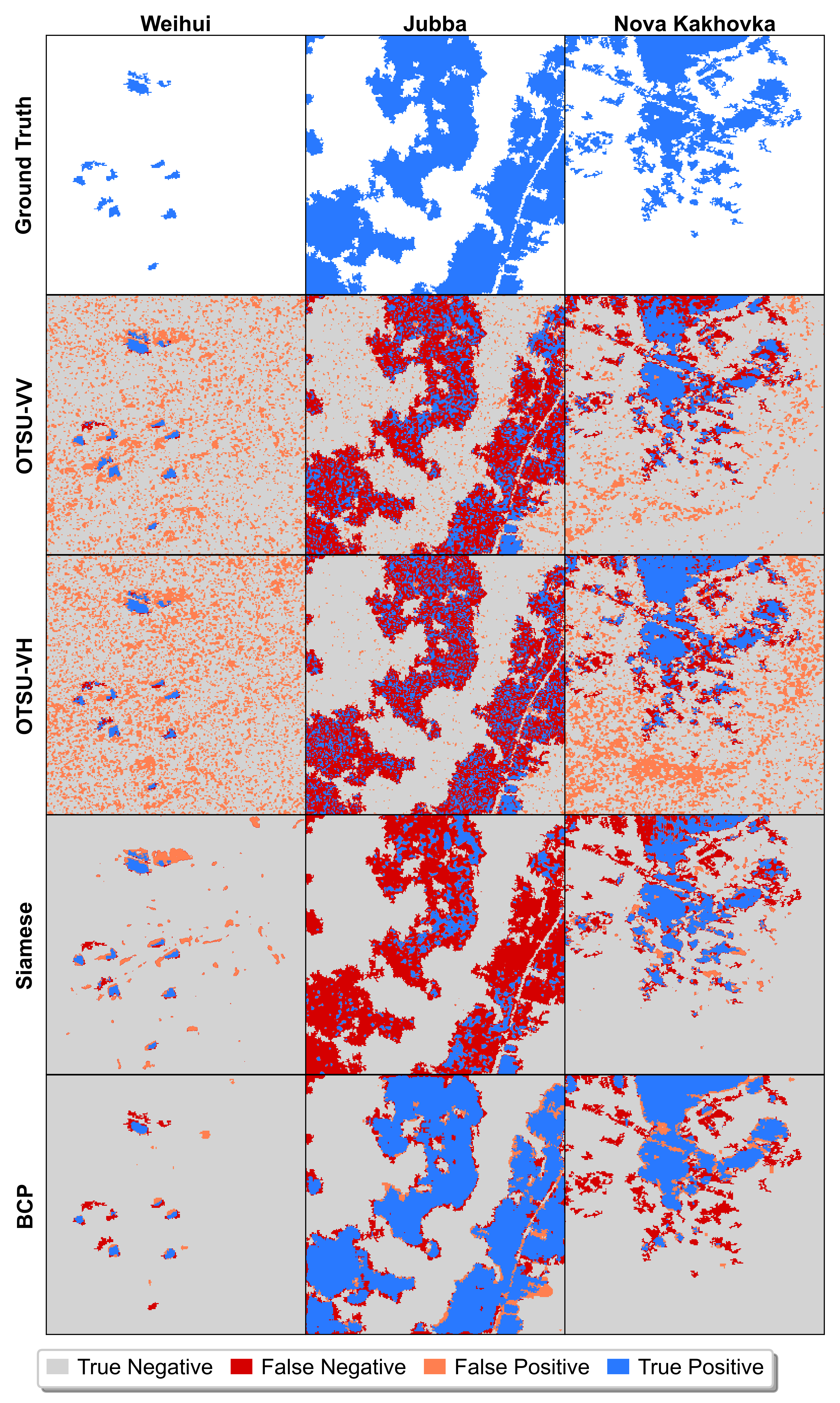}
\caption{Comparative flood detection results across three example patches in Weihui (China), Jubba (Somalia), and Nova Kakhovka (Ukraine) using different methods. Rows show Ground Truth (UrbanSARFloods reference), Otsu's thresholding (VV and VH polarizations), Siamese CNN, and the proposed BCP method.}
\label{fig:fig6}
\end{figure*}

With the use of the optimal parameters for BCP, Figure \ref{fig:fig6} illustrates the comparison of our method with the conventional and deep-learning approaches across randomly selected three example patches.
Each column represents a different study area, with rows showing the referenced Ground Truth data, Otsu's thresholding (VV, VH), Siamese CNN, and BCP methods. The Otsu's thresholding approaches show poor performance across all sites, with significant false positives (overestimation) in Weihui and substantial false negatives (underestimation) in Jubba and Nova Kakhovka. Between the two Otsu's thresholding variants, VV input appears to perform slightly better than VH input in terms of reducing scattered noise, though both struggle with accuracy. The Siamese CNN demonstrates considerably improved performance with reduced noise compared to the Otsu's thresholding approaches, successfully eliminating much of the scattered false positives. However, it tends to underestimate the flood extent, particularly evident in Jubba and Nova Kakhovka where significant false negatives (red areas) are present. The BCP method shows the most promising results, achieving good flood detection with minimal noise while better capturing the full flood extent across all three study areas.

\subsection{Quantitative Performance Evaluation}

The table \ref{tab:comparison} summarizes the accuracy measures of our approach compared to Otsu's thresholding and Siamese CNN approaches. For this evaluation, we used the globally optimized parameters identified through sensitivity analysis ($t=0.2$, $w=9$) with VV-VH input.
Our BCP approach achieved F1 scores ranging from 0.49 to 0.75, significantly outperforming Otsu's thresholding and Siamese CNN approaches across all sites.
Otsu's thresholding approach struggled with false positives, resulting in low precision across all sites. While the Siamese CNN showed improved performance over Otsu's thresholding approach, it still fell short of our method's capabilities.

\begin{table*}[!t]
\caption{Flood detection accuracy for Otsu's thresholding, Siamese CNN, and Bayesian analysis of change point problems (BCP), evaluated across three flood events (Weihui, Jubba, and Nova Kakhovka), with the highest value in each column shown in bold.}
\label{tab:comparison}
\centering
{\scriptsize
\begin{tabular}{|l|cccc|cccc|cccc|}
\hline
& \multicolumn{4}{c|}{Weihui} & \multicolumn{4}{c|}{Jubba} & \multicolumn{4}{c|}{Nova Kakhovka} \\
& Precision & Recall & F1 & IoU & Precision & Recall & F1 & IoU & Precision & Recall & F1 & IoU  \\
\hline
Otsu (VV) & 0.01 & 0.35 & 0.03 & 0.01 & 0.07 & 0.54 & 0.12 & 0.07 & 0.02 & \textbf{0.58} & 0.04 & 0.02 \\
Otsu (VH) & 0.02 & 0.56 & 0.05 & 0.02 & 0.08 & 0.61 & 0.14 & 0.07 & 0.02 & 0.51 & 0.04 & 0.02 \\
Siamese-CNN & \textbf{0.72} & 0.23 & 0.35 & 0.21 & \textbf{0.90} & 0.26 & 0.40 & 0.25 & \textbf{0.72} & 0.23 & 0.35 & 0.21 \\
BCP & 0.71 & \textbf{0.78} & \textbf{0.75} & \textbf{0.59} & 0.60 & \textbf{0.78} & \textbf{0.68} & \textbf{0.52} & 0.52 & 0.46 & \textbf{0.49} & \textbf{0.36} \\
\hline
\end{tabular}
}
\end{table*}

\subsection{Performance Variations Between Urban and Open Areas}

To better understand the performance characteristics across different land cover types, we conducted separate analyses for flood urban and flood open pixels. Tables \ref{tab:FO} and \ref{tab:FU} present the detailed accuracy metrics for each category and method.
In open areas, all methods demonstrated measurable effectiveness in flood detection, though with varying degrees of success. The BCP method achieved the highest performance in Weihui with a precision of 0.71 and a recall of 0.78 (IoU: 0.59), followed by results in Jubba (precision: 0.60, recall: 0.78, IoU: 0.52). However, its performance declined significantly in Nova Kakhovka (precision: 0.59, recall: 0.46, IoU: 0.36). The Siamese CNN demonstrated a modest performance in open areas, particularly in Weihui where it achieved an IoU of 0.46, though its effectiveness varied considerably across sites. The Otsu's thresholding approach showed a poor performance across sites, with recall values ranging from 0.33 to 0.61, though precision remained low (0.01-0.08). 

The analysis of urban environments revealed more challenging conditions for flood detection across all methods. The conventional Otsu's thresholding approach exhibited extremely low precision (0.00) despite achieving moderate recall values (0.63-0.88), resulting in poor overall performance with F1 scores (0.00). Similarly, the Siamese CNN demonstrated a poor performance in urban areas, with an F1 score and IoU of zero. The BCP method, while excelling in open areas, showed limited effectiveness in urban environments with precision below 0.05 and recall under 0.16 across all sites.

\begin{table*}[!t]
    \caption{Flood detection accuracy for Otsu's thresholding, Siamese CNN, and Bayesian analysis of change point problems (BCP), evaluated across three flood events (Weihui, Jubba, and Nova Kakhovka) for open areas, with the highest value in each column shown in bold.}
    \label{tab:FO}
    \centering
    {\scriptsize
    \begin{tabular}{|l|cccc|cccc|cccc|}
    \hline
    & \multicolumn{4}{c|}{Weihui} & \multicolumn{4}{c|}{Jubba} & \multicolumn{4}{c|}{Nova Kakhovka} \\
    & Precision  & Recall  & F1  & IoU  & Precision  & Recall  & F1  & IoU  & Precision  & Recall  & F1  & IoU  \\
    \hline
    Otsu (VV) & 0.01 & 0.33 & 0.02 & 0.01 & 0.07 & 0.54 & 0.12 & 0.07 & 0.02 & \textbf{0.58} & 0.04 & 0.02 \\
    Otsu (VH) & 0.02 & 0.56 & 0.04 & 0.02 & 0.08 & 0.61 & 0.14 & 0.07 & 0.02 & 0.51 & 0.04 & 0.02 \\
    Siamese-CNN & 0.63 & 0.62 & 0.63 & 0.46 & 0.57 & 0.26 & 0.36 & 0.22 & 0.21 & 0.15 & 0.17 & 0.09 \\
    BCP & \textbf{0.71} & \textbf{0.78} & \textbf{0.74} & \textbf{0.59} & \textbf{0.60} & \textbf{0.78} & \textbf{0.68} & \textbf{0.52} & \textbf{0.59} & 0.46 & \textbf{0.49} & \textbf{0.36} \\
    \hline
    \end{tabular}
    }
\end{table*}

\begin{table*}[!t]
        \caption{Flood detection accuracy for Otsu's thresholding, Siamese CNN, and Bayesian analysis of change point problems (BCP), evaluated across three flood events (Weihui, Jubba, and Nova Kakhovka) for urban areas, with the highest value in each column shown in bold.}
        \label{tab:FU}
        \centering
        {\scriptsize
        \begin{tabular}{|l|cccc|cccc|cccc|}
        \hline
        & \multicolumn{4}{c|}{Weihui} & \multicolumn{4}{c|}{Jubba} & \multicolumn{4}{c|}{Nova Kakhovka} \\
        & Precision  & Recall  & F1  & IoU  & Precision  & Recall  & F1  & IoU  & Precision  & Recall  & F1  & IoU  \\
        \hline
        Otsu (VV) & 0.00 & \textbf{0.70} & 0.00 & 0.00 & 0.00 & \textbf{0.88} & 0.00 & 0.00 & 0.00 & 0.69 & 0.00 & 0.00 \\
        Otsu (VH) & 0.00 & 0.63 & 0.00 & 0.00 & 0.00 & \textbf{0.88} & 0.00 & 0.00 & 0.00 & \textbf{0.72} & 0.00 & 0.00 \\
        Siamese-CNN & 0.00 & 0.01 & 0.00 & 0.00 & 0.00 & 0.00 & 0.00 & 0.00 & 0.00 & 0.02 & 0.00 & 0.00 \\
        BCP & \textbf{0.05} & 0.13 & \textbf{0.07} & \textbf{0.03} & 0.00 & 0.03 & 0.00 & 0.00 & \textbf{0.03} & 0.16 & \textbf{0.05} & \textbf{0.03} \\
        \hline
        \end{tabular}
        }
\end{table*}

\section{Discussion}
\label{sec:Discussion}

\subsection{Methodological Advances in SAR-Based Flood Detection}

Our study introduces Bayesian analysis for change point problems (BCP) to flood monitoring applied to Sentinel-1 GRD time series data. Unlike supervised methods that require extensive labeled training data and face challenges with geographical transferability \cite{Wu2023-kc,Ghosh2024-ub}, our approach leverages the statistical properties of SAR time series to detect flood-induced changes without any training requirements. This fundamental training-free characteristic is a substantial advantage, as it completely eliminates the overfitting issues inherent in deep learning and other machine learning approaches. Machine learning methods often perform well in areas represented in training datasets but fail when applied to novel geographical contexts or unusual flood patterns \cite{Amitrano2024-av, Zhao2025-cx}. By contrast, our BCP method requires no supervised learning process, making it immediately applicable to any region with sufficient SAR time series data, regardless of whether similar flood events have been previously observed and labeled.

Our method advances the field by reframing flood detection as a temporal changepoint identification problem rather than a binary classification task. Conventional classification approaches typically rely on absolute backscatter values or simple change metrics between two images, making them vulnerable to variations in instrument calibration, incidence angle effects, and seasonal changes in land cover \cite{Schlaffer2015-uv}. In contrast, our Bayesian framework models the entire temporal trajectory of backscatter values, enabling more robust discrimination between normal fluctuations and flood-induced changes.

Parameter sensitivity analysis showed that a window size of 9 pixels provided a good balance between noise reduction and preservation of flood boundary details, though our method demonstrated relatively stable performance across a range of parameter settings. 
Unlike most previous studies that applied fixed parameter settings for speckle suppression in SAR imagery, we explored various parameter combinations to maximize the effectiveness of BCP outputs. Our analysis revealed that $w=9$ and $t=0.2$ yielded the maximum F1 score for our studies (Figure \ref{fig:sensitivity}). However, it should be noted that these optimal parameters were determined using the test dataset with a spatial resolution of 20-m provided by UrbanSARFloods. Optimal parameter settings should be determined for the original Sentinel-1 GRD data with a spatial resolution of 10-m for real-world applications.

\subsection{Performance in Relation to State-of-the-Art Approaches}

Our performance analysis reveals distinct patterns across different environmental contexts and detection methods. In open areas (Table \ref{tab:FO}), our BCP method demonstrated robust performance with F1 scores ranging from 0.49 to 0.74, outperforming Otsu's thresholding approach (F1: 0.02-0.14). The Siamese CNN showed variable performance in open areas, achieving its best results in Weihui (F1: 0.63) but performing less effectively in Jubba (F1: 0.36) and Nova Kakhovka (F1: 0.17). These results suggest that while deep learning approaches can achieve moderate success, the model may suffer from insufficient training to identify flood patterns in unknown landscape characteristics.

The stark performance disparity between open and urban areas can be attributed to fundamental differences in SAR backscatter characteristics \cite{Zhao2025-cx}. Open areas offer more uniform backscatter patterns in SAR imagery, enabling more reliable flood detection across all methods due to clearer contrast between flooded and non-flooded surfaces. In contrast, urban environments present complex scattering mechanisms due to built infrastructure, leading to increased signal ambiguity \cite{Amitrano2024-av, Zhao2025-cx}. The presence of various materials and structures in urban areas creates multiple reflection pathways, making flood detection particularly challenging. These physical constraints fundamentally limit the effectiveness of current detection methods, regardless of their computational sophistication.

\subsection{Integration with Operational Flood Monitoring Systems}

The practical application of our method extends beyond academic research to operational flood monitoring systems. Our approach could complement existing near-real-time flood mapping services such as the Copernicus Emergency Management Service (EMS) \cite{copernicus2025} and NASA's MODerate Resolution Imaging Spectroradiometer (MODIS)/Visible Infrared Imaging Radiometer Suite (VIIRS) near-real-time global flood products \cite{nasa_modis_flood_2025, nasa_viirs_daily_flood_2025, nasa_viirs_hourly_flood_2025}. 
These systems currently rely primarily on optical imagery or simple thresholding of SAR data, facing challenges with cloud cover and discrimination between permanent and temporary water features.

For regional flood monitoring centers with limited computational resources, our method offers particular advantages. Unlike deep learning approaches that typically require GPU acceleration and substantial training data, our statistical framework operates efficiently on standard computing hardware and requires no training phase. This accessibility could enable more widespread adoption of SAR-based flood monitoring in developing regions where flood impacts are often most severe but technical resources are limited.

The method's ability to process original Sentinel-1 GRD data directly also streamlines the operational workflow, reducing the need for extensive preprocessing chains that can introduce delays in time-sensitive emergency response scenarios. Combined with Sentinel-1's systematic acquisition plan, this could enable near-real-time flood alerts within hours of satellite overpass, a critical improvement over current systems that often require days to deliver actionable information.

\subsection{Limitations and Future Research Directions}

Despite the promising results, several limitations of our approach warrant consideration. The method's performance in open areas, particularly in Nova Kakhovka, indicates challenges in distinguishing flood inundation from other changes in agricultural regions. This limitation appears related to the complex temporal patterns of vegetation growth and harvest activities that can create backscatter changes similar to flooding \cite{Amitrano2024-av}. 
Additionally, our approach often misclassify non-flood changes in SAR backscatter as floods due to complex scattering mechanisms that impair water detection, particularly in urban environments. \cite{Amitrano2024-av, Zhao2025-cx}. Future research could address these limitations by developing more sophisticated time series analysis techniques that can better discriminate between different types of change patterns within the SAR signal itself, enhancing the algorithm's ability to identify unique flood signatures without relying on external datasets.

From a methodological perspective, the current BCP implementation assumes normally distributed observation values \cite{Barry1993-qp}, which may not fully capture the complex statistical properties of SAR data. Alternative Bayesian frameworks that incorporate more sophisticated distributions, such as the gamma distribution commonly used for SAR intensity modeling, could potentially improve detection accuracy. Similarly, hierarchical Bayesian models that explicitly account for spatial context could enhance the method's robustness to speckle without requiring post-processing spatial filtering. These approaches are specifically designed to address the challenge of flood detection in urban areas. 

The temporal resolution of Sentinel-1 (6-12 days) represents another limitation for monitoring rapidly evolving flood events \cite{Munasinghe2023-oc}. Future research could explore the integration of our approach with constellation SAR observations to achieve higher temporal frequency. Alternatively, data fusion techniques that combine SAR with optical or passive microwave observations could provide more continuous monitoring capabilities, leveraging the complementary strengths of different sensor types.

\section{Conclusions}
\label{sec:Conclusions}

We present a novel training-free approach for automated flood detection using Bayesian analysis for change point problems (BCP) applied to Sentinel-1 GRD time series data. Our method analyzes one-year baseline periods of SAR backscatter intensity to statistically model normal temporal behavior, then computes posterior probabilities of change points at flood observation dates. Using combined VV and VH polarizations with optimized post-processing parameters (moving-window kernel size = 9, threshold = 0.2), the approach was validated across three diverse geographical contexts. 

The BCP method substantially outperformed baseline approaches, particularly when compared to the Siamese CNN deep learning technique, achieving F1 scores 140–214\% higher. A key benefit of our approach is that it requires no training data, unlike deep learning methods. Across all test sites, BCP consistently outperformed baseline methods, with F1 scores of 0.75 (Weihui), 0.68 (Jubba), and 0.49 (Nova Kakhovka).

The training-free nature of our approach eliminates overfitting issues and enables immediate deployment to any region with sufficient SAR coverage, regardless of whether similar flood events have been previously observed. This accessibility is particularly valuable for disaster response in data-scarce regions where technical resources are limited. Our method's ability to process original Sentinel-1 GRD data directly streamlines operational workflows, potentially enabling near-real-time flood alerts within hours of satellite overpass.
However, the 6-12 day temporal resolution of Sentinel-1 remains a constraint for monitoring rapidly evolving flood events. Future research should explore integration with higher-frequency SAR constellations and investigate more sophisticated Bayesian frameworks that better capture the complex statistical properties of SAR data, particularly for improving urban flood detection where current performance remains challenging.

\section*{Acknowledgments}
\label{sec:Acknowledgements}
This work was funded by the Disaster
Prevention Research Institute’s Implementation Science Research for Regional Communities (Specific) project at  Kyoto University (Project No. 2024RS-01) and NEDO (23200859-0).

 \bibliographystyle{unsrt} 
 \bibliography{cas-refs}

\end{document}